# Characterisation of SATCOM Networks for Rapid Message Delivery: Early In-Orbit Results.

Robert Mearns, Airlie Chapman, Michele Trenti
The University of Melbourne, Melbourne Space Lab
Grattan Street, Parkville, Victoria 2010, Melbourne, Australia;
robert.mearns@unimelb.edu.au

## ABSTRACT

Traditional nanosatellite communication links rely on infrequent ground-station access windows. While this is well suited to both payload data and detailed scheduling information, it leads to both intermittent and short communication windows. The resulting long periods without contact are ill-suited for both opportunistic tasking of satellites and triggers generated by autonomous operations. Existing orbital infrastructure in the form of satellite communication (SATCOM) networks, such as Iridium and others provide a readily available and cost effective solution to this problem. While these networks continue to be utilized onboard nanosatellites, a full characterization of their utility and performance in-orbit is vital to understand the reliability and potential for high-timeliness message delivery. The Space Industry Responsive Intelligent Thermal (SpIRIT) 6U nanosatellite is a mission led by The University of Melbourne in cooperation with the Italian Space Agency. SpIRIT received support from the Australian Space Agency and includes contributions from Australian space industry and international research organizations. Developed over the last four years and launched in a 510km Polar Sun Synchronous Orbit in late 2023, SpIRIT carries multiple subsystems for scientific and technology demonstration. The Mercury subsystem provides a demonstration and characterization test bed for the features and capabilities of autonomous SATCOM utilization in-orbit, while also providing the capability of rapid down-link of detection events generated by the main scientific payload of the mission, the HERMES instrument, a gamma and x-ray detector for the detection of high-energy astrophysics transients (Gamma Ray Bursts). This paper first presents a brief payload overview and overview of the experimental design of the characterization efforts. From this, early in-orbit results are presented along with a comparison to ground-based experiments, focusing on lessons learned throughout the mission development and operations. This work not only sheds light on the utility of these networks for autonomous operations, and on their potential impact to enable greater utilization of nanosatellites for scientific missions, but also offers insights into the practical challenges related to the design and implementation of utilizing these networks in-orbit.

## INTRODUCTION

The Mercury payload, developed by the Melbourne Space Laboratory at the University of Melbourne, Australia was designed and fabricated for the SpIRIT nanosatellite mission. This mission, funded by the Australian Space Agency, is the first to carry a foreign agency payload – the HERMES instrument, an X-ray and γ-ray detector provided by the Italian Space Agency for the detection of high-energy astrophysical transients such as Gamma Ray Bursts (GRBs). SpIRIT and the constellation of which it will ultimately be a part, the HERMES scientific pathfinder [1], will yield sky localizations of GRBs via cross-correlations of time-stamped photon counts between elements of the constellation, providing a localization of the event in the sky using trilateration between measurements of the arriving wave-front. As the correlation process requires information from multiple elements of the constellation, correlation can only occur once event data has been transmitted to the ground from multiple constellation elements. Furthermore, as these events fade rapidly with time, it is preferable to produce a localization as rapidly as possible. For a moderately sized ground-station network, the delay introduced by the time to the next ground station pass for each element becomes untenable.

Thus, for events generated by autonomous operations, such as the detection of GRBs by the HERMES pathfinder constellation, traditional nanosatellite communication links provided by dedicated ground-stations are ill-suited for the timely delivery of these events. Similarly, detection of ground-based events such as bush-fires or other natural disasters, [2] by small, agile, and importantly inexpensive, orbital observation spacecraft would be well served by the ability to rapidly deliver unsolicited triggers generated by the observing satellite to users on the ground.

In the opposite communication direction, the ability to rapidly task a satellite to take advantage of detected events for further follow up [3], would benefit from data relay capabilities without the need to wait for a ground-station pass.

These high-timeliness capabilities are provided to large missions via the Tracking and Data Relay Satellite (TDRS) system in the US, and the European Data Relay Satellite (EDRS) system in Europe. However, these networks are oversubscribed, extremely expensive, and not available to small organizations or missions.

Existing and planned commercial orbital communication infrastructure has been proposed as a viable alternative to ground-station networks for rapid up-link and down-link. For example, NASA's Communication Services Project (CSP) is funding the exploration of commercial data relay services [4], with some recent successes by some of the recipients[5]. However, these services are likely many years off.

Separately to government supported endeavors, commercial satellite communication modems have been employed on a number of small satellites within the nanosatellite community to support operational activities [6], [7] with at least one commercial company (Near Space Launch) providing a COTS application of a SATCOM modem [8]. Although [8] reports great success with their value-add nanosatellite COTS modem for the Iridium network, the underlying model applicable when adapting an existing SATCOM network intended for ground usage to orbit has not been widely explored in a quantitative manner in public literature.

## SATCOM CONSTELLATION UTILIZATION IN ORBIT

Applications of existing orbital SATCOM infrastructure to nanosatellite communications have been explored, using both the Globalstar constellation, [9], [10], [6], and [11], as well as the Iridium constellation, [12], [13], [7], [8]. While additional networks exist, and several new networks are proposed, the Iridium network remains the only network showing promise for use on-board small platforms such as nanosatellites in the near future. Until recently Globalstar was also an option, however the company has ceased supporting orbital modems, leaving Iridium as the only extant option at the moment.

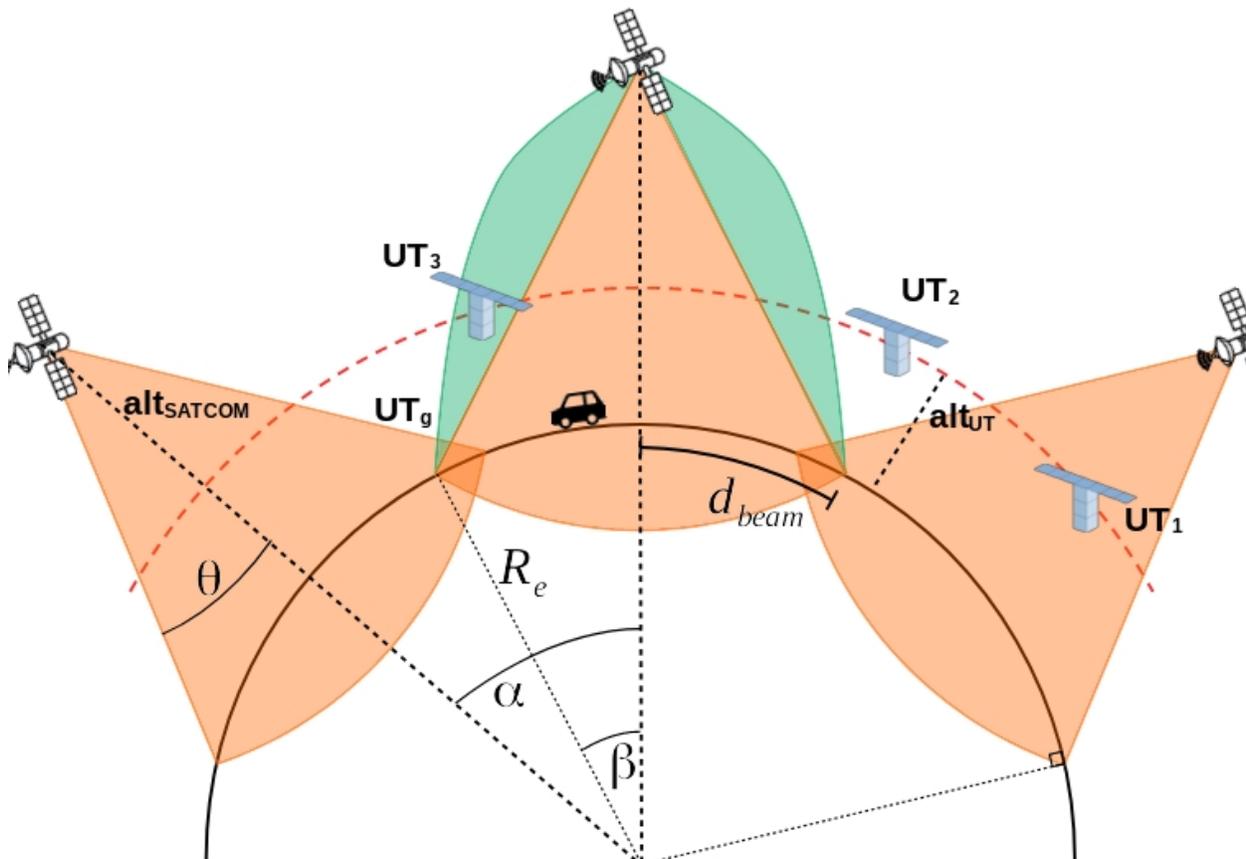

**Figure 1: A network constellation at altitude $alt_{SATCOM}$ provides communication beams, in orange, designed to overlap for a ground based User Terminal ($UT_G$), but leaves gaps in coverage for a user terminal at orbital altitudes ($UT_2$).**

Table 1: SATCOM Constellation Traversal Times for Orbital UT Altitudes

| Constellation | 200km [s] | 300km [s] | 400km [s] | 500km [s] | 600km [s] | 700km [s] | 800km [s] |
|---|---|---|---|---|---|---|---|
| Globalstar | 286 | 253 | 223 | 196 | 171 | 147 | 125 |
| Iridium NEXT | 174 | 137 | 104 | 75 | 47 | 20 | - |
| Orbcomm | 146 | 112 | 82 | 53 | 26 | 1 | - |
| Inmarsat | 958 | 939 | 925 | 913 | 904 | 896 | 889 |
| Thuraya | 958 | 939 | 925 | 913 | 904 | 896 | 889 |
| OneWeb | 61 | 55 | 49 | 43 | 37 | 31 | 25 |
| Starlink | 86 | 60 | 35 | 12 | - | - | - |

Since SATCOM constellation networks are designed for ground users, their coverage figures are stated in terms of ground spot size, for example, the Iridium network is quoted as having global coverage. However at orbital altitudes, the coverage of the network will be reduced as a function of the radiation pattern of the network satellites. This results in significant coverage gaps at typical nanosatellite orbital altitudes, with coverage gaps growing with altitude, see Figure 1.

Prior simulation works [9],[11],[12],[14],[15] have assumed a simplified model of network communication beams using a conical beam defined using the ground spot size (orange beams in Figure 1). The apex angle of the conical beam is determined from the diameter of the circle approximating the -3dB (FWHP) ground spot. The cone sizes of various constellations under this assumption are given in Table 2, derived from declared information in communication filings for each constellation.

It should be noted that the OneWeb and Starlink constellations do not currently support 'short-burst' machine to machine transmission services which are well suited to the intermittent contact windows available to an orbiting network modem, referred to as a User Terminal (UT). Furthermore, as these networks have electrically steered phased arrays, as opposed to fixed cells based on GSM technology, it is not clear that they are suitable to support intermittently connected, fast moving UTs. They have been included in Table 2 and Table 1 for reference only.

As an extension to the conical model of the beam, [16]

Table 2: SATCOM Constellation Altitude and Conical Beam Apex Angles

| Constellation | Altitude [km] | Apex Angle [°] |
|---|---|---|
| Globalstar | 1,414 | 108 |
| Iridium NEXT | 780 | 125.8 |
| Orbcomm | 703 | 124 |
| Inmarsat | 35,786 | 17.2 |
| Thuraya | 35,786 | 17.2 |
| OneWeb | 1200 | 50 |
| Starlink | 550 | 120 |

theorized that the shorter path length to an orbital UT would compensate for the reduction in signal strength further off-axis, resulting in an effective increase in the diameter of the spot at that altitude over the conical model; $UT_3$ will retain signal connection with the network while it is the green region, Figure 1.

Under either model of a constellation beam, the traversals of an orbiting UT through a constellation beam will be relatively short, presenting limited opportunity for the UT to complete the Acquisition, Access, Registration and Telephony processes required to transmit a message via these networks. Further compounding this effect will be any position dependent changes to the network channel parameters as the UT traverses the network beam; frequency re-use schemes employed across different cells within the ground-spot, timing frame assignment to different cells, etc.

Assuming a conical model of the beam, the diametric traversal times for varying altitudes are given in Table 1. These times represent the maximum possible traversal time a UT would experience as it travels along a diametric chord, for a given altitude. Critically, these times are calculated for UT traversals perpendicular to the network satellite. While prograde or retrograde motion will of course have a large impact on these figures, they nevertheless provide a baseline for estimates of potential coverage times over the course of an orbit.

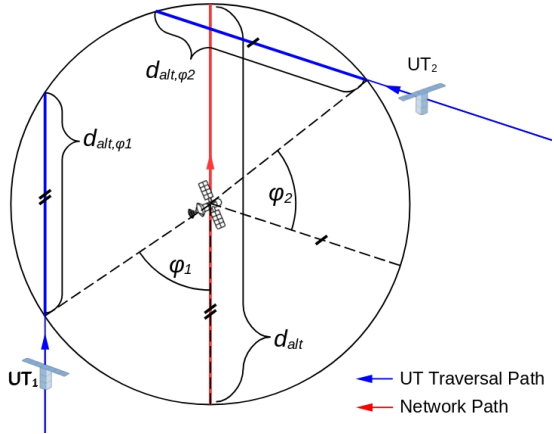

**Figure 2: Traversal chord for a given altitude and entering angle $\varphi$, given by $d_{alt,\varphi}$. Entering angle taken from the parallel beam diameter.**

Traversal times for orbital paths further from the SATCOM satellite nadir will be given by the chord to diameter fraction of the diametric traversal time. Figure 2 illustrates traversal chords as a function of both the orbital altitude and $\varphi$, the angle of the entry point from the parallel diameter. For $UT_2$ in Figure 2 the traversal chord will not be a straight line, as the SATCOM satellite is moving at comparable speeds, consequently in the frame of the SATCOM satellite, the traversal chords of all UTs will be curved towards the wake direction of the network satellite. However, this effect has been ignored for the first order analysis considered here.

In order for a short-burst 'call' with a GSM technology based network, such as Iridium or Globalstar, to successfully complete, the UT must progress through the Acquisition, Access, Registration and Telephony processes. For a short-burst transmission, in which the data packets are only up to several hundred bytes, the 'call' time is dominated by the Acquisition, Access and Registration processes, while the data packet transmission time is limited to a few hundred milliseconds.

Grouping the Acquisition, Access and Registration processes into $t_{access}$; the time taken for a single packet transmission, without application layer acknowledgment, is given as (1) for N network satellites participating in the relay to the ground [17].

$$t_{packet} = t_{access} + t_{uplink} + (N-1)t_{crosslink} + (N)t_{sat\_processing} + t_{downlink} \quad (1)$$

The time taken to perform a search through the parameter space defining the channel reuse schemes, $t_{access}$, presents the greatest unknown for call timing. Figure 3 compiled by [16] (and reproduced here) indicates that the mean time for an Iridium 'short-burst' data session to complete on the ground is ~6secs with a long tail out to ~25secs.

As this distribution represents the time taken to search through the channel parameters, if the underlying parameter space is modified by conditions experienced in orbit, the distribution will likely change. One highly likely scenario is the doppler shift experienced due to the relative velocities of the UT and networks satellite affecting the spacing between frequency re-use channels, and therefore changing the search results.

While the UTs considered here can compensate for doppler shifts close to what would be experienced in orbit, it is possible that the distributions shown in Figure 3 will therefore stretch to the right when accounting for orbital conditions.

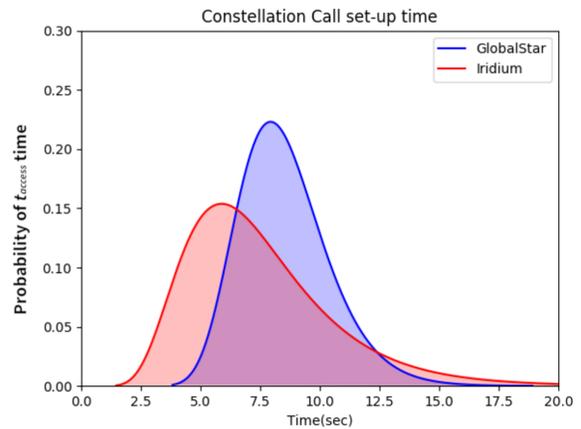

**Figure 3: Probability of $t_{access}$ times for the Globalstar and Iridium constellations for a ground UT, reproduced from [16] with permission.**

Figure 4 shows the $t_{access}$ distribution relative to the traversal times of the entire Iridium beam for varying altitude and entry angles. While for altitudes less than ~650km it appears that a traversal time is significantly longer than the $t_{access}$, there are two important caveats.

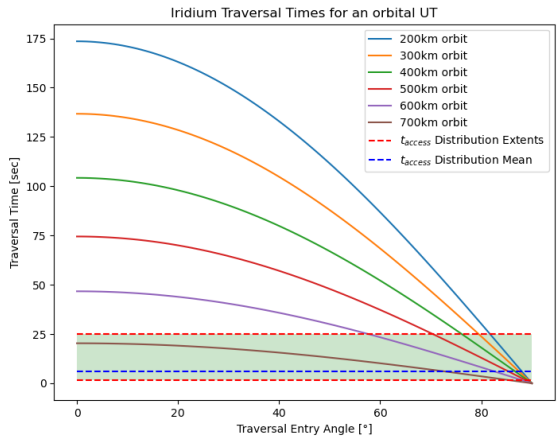

**Figure 4: Traversal times of an orbital UT for varying traversal entry angles, as compared to baseline call setup times.**

1. The relative frequency distribution of traversal entry angles will not be uniform for any given UT orbit, and is more likely to be weighted towards larger entry angles, meaning that shorter traversal times are more likely.

2. $t_{access}$ includes the time duration for: pilot searches, channel frequency and timing adjustments, and synchronization. For any of these processes which depend on searching for valid parameters, the process in unlikely to succeed if the parameter changes during the process. Thus there is a higher likelihood of successful access to the network if the $t_{access}$ duration is sufficiently short to ensure the underlying channel parameters remain static during $t_{access}$. Depending on the frequency and timing frame re-use schemes between cell spots within a SATCOM beam, these parameters may only remain static for a single cell spot (this will generally be the case, as re-use schemes are intended to minimize interference between adjacent cells), significantly reducing the length of periods during a traversal in-which $t_{access}$ may complete.

In the analysis framework outlined above, satellite beams are assumed to be homogeneous, with channel parameters remaining static throughout the entire beam, (and hence traversal). As eluded to above, this will almost certainly not be the case, suggesting that the above framework may more accurately reflect traversals through sub-cells within the beam, rather than the entire beam.

Attempting to communicate via these networks without consideration of both the likelihood of instantaneous network coverage at that location and time, and continued network coverage for the duration of $t_{packet}$, will however, certainly result in many failed communication attempts. Since any communication attempt with these networks is associated with both resource utilization (power, attitude budget, etc.) and operational opportunity costs (time off target to facilitate communication attempts), the total number of failed communication attempts should be minimized.

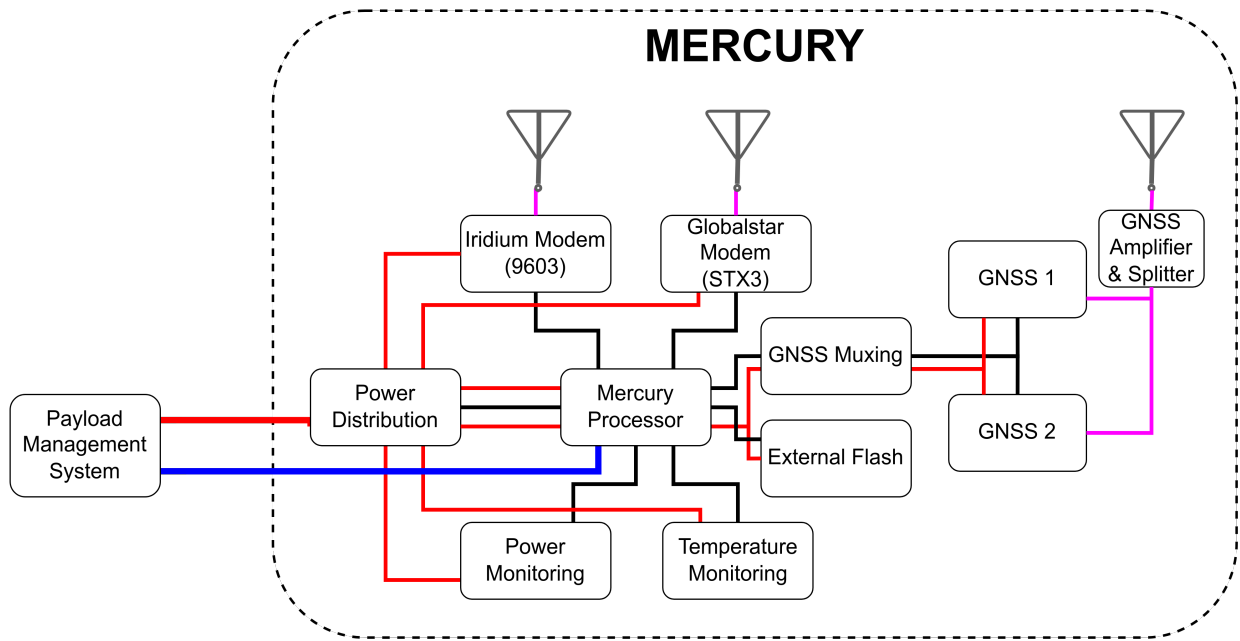

**Figure 5: Simplified block diagram of the Mercury payload. Power links shown in red; data links shown in black; RF links shown in magenta.**

Therefore, a thorough understanding of those network parameters which are likely to affect an orbital UT will allow for the design of policies which govern communication attempts, thereby optimizing resource utilization and operational opportunity.

**MERCURY HARDWARE CONCEPT**

The Mercury payload aboard the SpIRIT nanosatellite is intended to:

1. Provide a test bed for the characterization of constellation parameters affecting the utilization of SATCOM constellations by orbital UTs

2. Provide high-timeliness down-link of potential GRB event data from the HERMES payload

3. Provide high precision timing to the HERMES instrument in order to discipline its onboard chip scale atomic clock (CSAC)

4. Provide lifetime monitoring of the health of each modem

To accomplish these objectives, Mercury is designed to use two Commercial Off The Shelf (COTS) SATCOM UTs, Iridium and Globalstar to carry out measurements of their respective constellations throughout the lifetime of the mission. Additionally, Mercury carries two redundant GNSS modules to independently ascertain its location in orbit, relative to the satellite communication constellations, and additionally to distribute a pulse per second (1PPS) signal to the HERMES instrument for use by the CSAC.

Figure 5 outlines a simplified electrical block diagram of the Mercury payload, while Figure 6 shows the Mercury flight model. An external controlling application, in this case the 'Payload Management System' supplies external power and communication. Each of the blocks shown in Figure 5 are discussed briefly below.

**Power Distribution:** The Mercury payload is provided primary 12V an 5V lines by the payload management system. These primary lines are split and down-converted into 12V, 5V and 3V3 supply lines for the various components within each block shown in Figure 5. Switching is accomplished using integrated soft-start load switches. Management of the primary 12V and 5V lines is the responsibility of the external Payload Management System.

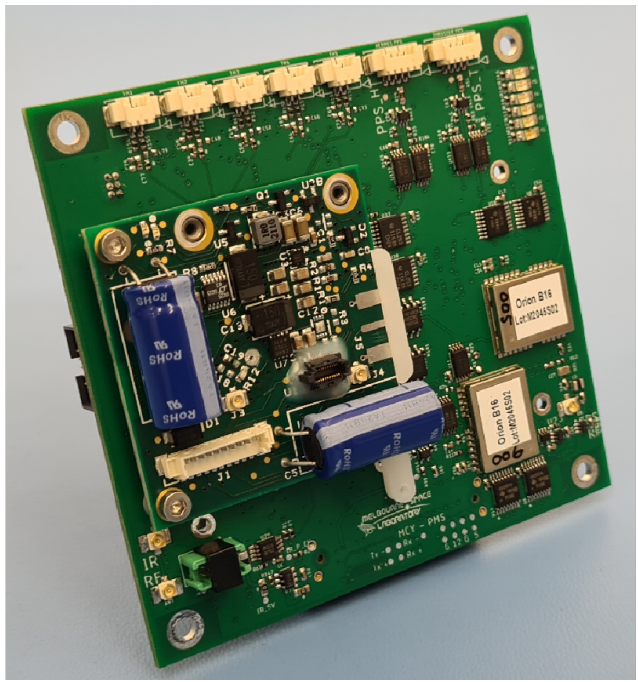
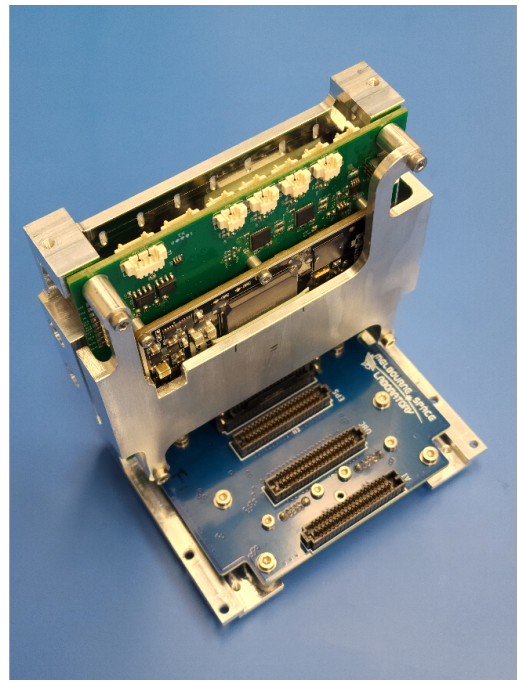

**Figure 6: Flight Model Mercury. Left - prior to integration (without Iridium Modem).**

**Right – During integration into the SpIRIT PMS Stack**

**Power Monitoring:** The current and voltage lines of each modem and GNSS are monitored throughout operations by a 12bit (11bit effective) ADC. Analog threshold monitoring is used to disable the modems or GNSSs if an over-current or under-voltage event is detected, with timestamped current and voltage logging used to determine any trends in power consumption over the lifetime of the mission.

**Temperature Monitoring:** The temperature of various critical locations associated with the Mercury payload are monitored to correlate any potential short or long term temperature dependent behavior.

Temperature measurement is accomplished using class A PT1000 RTDs configured in a ratiometric topology with a 16bit ADC. Additionally, 2-point calibration is provided to each ADC using external precision (±0.1%) resistors.

The areas of interest for temperature are outlined below.

- Underside of Globalstar antenna PCB
- GNSS amplifier RF can (located on the underside of the GNSS antenna)
- Edge of Iridium modem RF can.
- Mercury processor heatsink
- Underside of Iridium antenna
- Between PCB mounted GNSS modules

Mercury also provides additional temperature measurement for other payloads on-board SpIRIT.

**GNSS Multiplexing:** As the provision of a high-precision 1PPS signal to the HERMES instrument was considered paramount for the localization capability of the instrument, it was decided early in the design process that two cold redundant strings of the GNSS module and accompanying circuitry would be employed. This necessitated muxing of: various control lines to the GNSS modules; UART output from the modules; and the 1PPS output from the modules. Furthermore Mercury is capable of switching the 1PPS output between three different endpoints onboard SpIRIT: the HERMES instrument; the Payload Management System; and an additional payload. The 1PPS muxing was designed to introduce minimal jitter and propagation delay into the 1PPS signal edges, and has been shown to introduce less than <10ns of jitter into the chain.

**GNSS 1 & 2:** The GNSS modules are critical to the Mercury payload, both from a position determination aspect, and a timing distribution aspect for the HERMES instrument. The Iridium network requires UTs to provide a location when contacting the network. This 'registration' procedure does not require high precision (±250km radius), but is necessary for UTs which change their position between registration events, as is the case for an orbital UT.

Since the location precision for the purposes of the Iridium network is low, the driving requirement behind the selection of the GNSS modules was the timing jitter requirement of the HERMES instrument; <30ns jitter on the 1PPS edge to 1$\sigma$.

The GNSS modules selected were the NavSpark OrionB16 modules, a successor to the Venus 838FLPx. The OrionB16 module is capable of producing <5ns 1$\sigma$ jitter on the 1PPS edge under laboratory conditions, while weighing only 1.6g and consuming < 140mW.

**GNSS Amplifier & Splitter:** Since room on an external spacecraft face was only available for a single GNSS antenna for the Mercury payload, it was necessary to share the RF signal between the two modules. A Wilkinson power divider and an active two stage GNSS antenna with front-end Saw filter were selected to accomplish the splitting and ensure sufficient link margin after the split.

As the Iridium and GNSS frequencies are very close, 1.615GHz for Iridium, 1.57-1.6GHz for the various constellation L1 GNSS frequencies, the capability to disable the GNSS Amplifier while the Iridium modem is transmitting was also included.

**Iridium Modem:** The COTS Iridium 9603 modem with a carrier board provided by RockBLOCK, Figure 7. This carrier board provides power smoothing via two 5F supercapacitors, consequently, high current draw during transmit attempts is spread over a longer period.

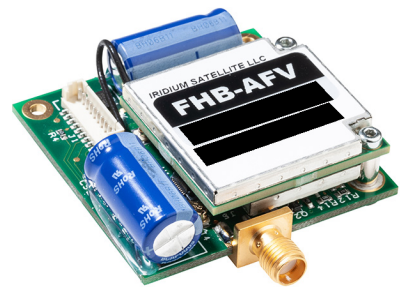

**Figure 7: Iridium 9603 modem with RockBLOCK carrier board.**

**Globalstar Modem:** The COTS Globalstar modem is a Simplex STX3 provided by NearSpaceLaunch, Figure 8.

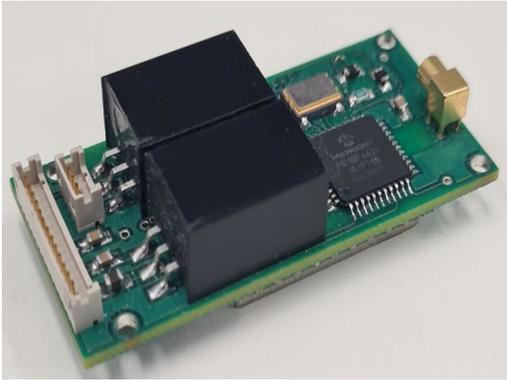

**Figure 8: Near Space Launch STX3 Globalstar Modem**

Unfortunately, Globalstar withdrew support of orbital modems late in the development of the SpIRIT mission. Consequently, both the modem and antenna are mounted in the SpIRIT satellite, but are electrically disconnected, preventing their use.

**Mercury Processor:** The Mercury processor is a COTS processor with flight heritage, supplied by Infinity Avionics. The Volkh processor is a daughterboard containing a SmartFusion2, external LPDDR, external MRAM, and latch-up protection. Although the Volkh processor provides reprogrammability functionality, this was not implemented on Mercury due to schedule pressure.

**External Memory:** Mercury carries 4Gb of external NAND flash with a custom data logging system implemented on the processor.

Each of these blocks are mounted to a single motherboard, either as populated components, or as a daughterboard.

The Mercury software architecture is designed to allow independent use of each of the Mercury subsystems by the controlling application, in this case the Payload Management System, in order to provide the necessary support functionality to the other payloads, as well as the potential for future in-orbit experimental design . Each subsystem has a controlling thread implemented as a FreeRTOS task, with commands issued to each thread by a dedicated 'dispatcher' thread.

SATCOM characterization experiments are then achieved by issuing commands from a dedicated 'experiment' thread to the 'dispatcher' thread, with each 'experiment' thread running a state machine of the experiment generally consisting of waiting for the necessary condition, executing the relevant modem actions, taking measurements, and then logging the outcome.

The purpose of each dedicated 'experiment' thread is detailed below. The experiment threads described below are applicable to the Iridium network. Equivalent experiments were implemented for the Globalstar network, but due to the withdrawal of support by Globalstar, they are considered defunct and not discussed.

It should be noted that the experiments and flow charts detailed below are representative of the as-flown state. A number of improvements have been considered since launch, drawing on the learnings gathered during both orbital and on-ground operations and experiments.

**Registration Timing:** Given the time taken to complete the acquisition/synchronization step of a call set-up, $t_{access}$, is assumed to be a stochastic process, repeated measurements of the time taken to complete a successful registration attempt with the Iridium network should yield a distribution bounding this process. The registration process consists of transmitting a data packet to the network in order to notify the network of the UT's location, the total time of a registration should be $t_{packet}$, with $t_{packet}$-$t_{access}$ a constant overhead on the stochastic process governed by the internal registration process of the Iridium telephony routing system .

Furthermore, by conducting the same experiment on the ground, any effects on the process due the orbital nature of the UT can be ascertained.

For each registration attempt, the following data is recorded: the result of the registration process with the network, the modem and antenna temperatures and the GNSS time and position the registration both began and ended. A flowchart of the experiment is shown in Figure 9.

**Message Transmission Timing:** Similar to the Registration Timing Experiment, the Message Transmission Timing experiment is intended to yield a distribution bounding the time taken by the modem to register and transmit a message. In this case however, the total time of a registration and transmission is the $t_{packet}$ under investigation. When compared to the registration attempt distribution, it is expected that there will merely be a constant time delay to account for the message packet transmission.

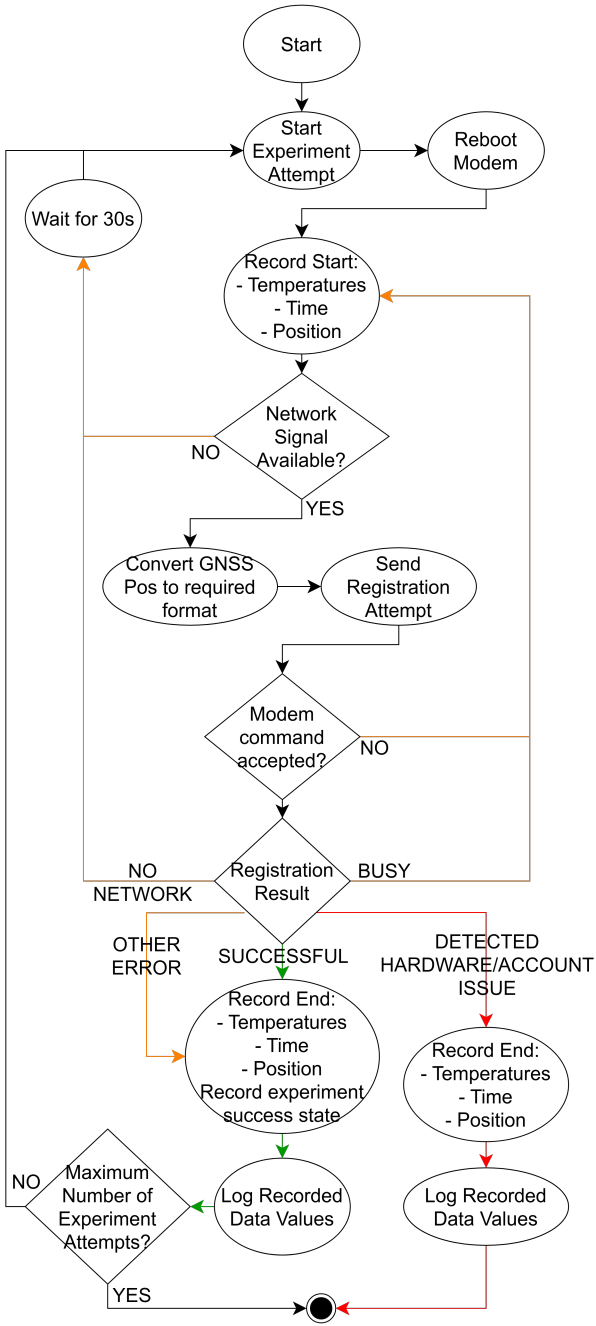

**Figure 9: Registration Timing Experiment Flowchart**

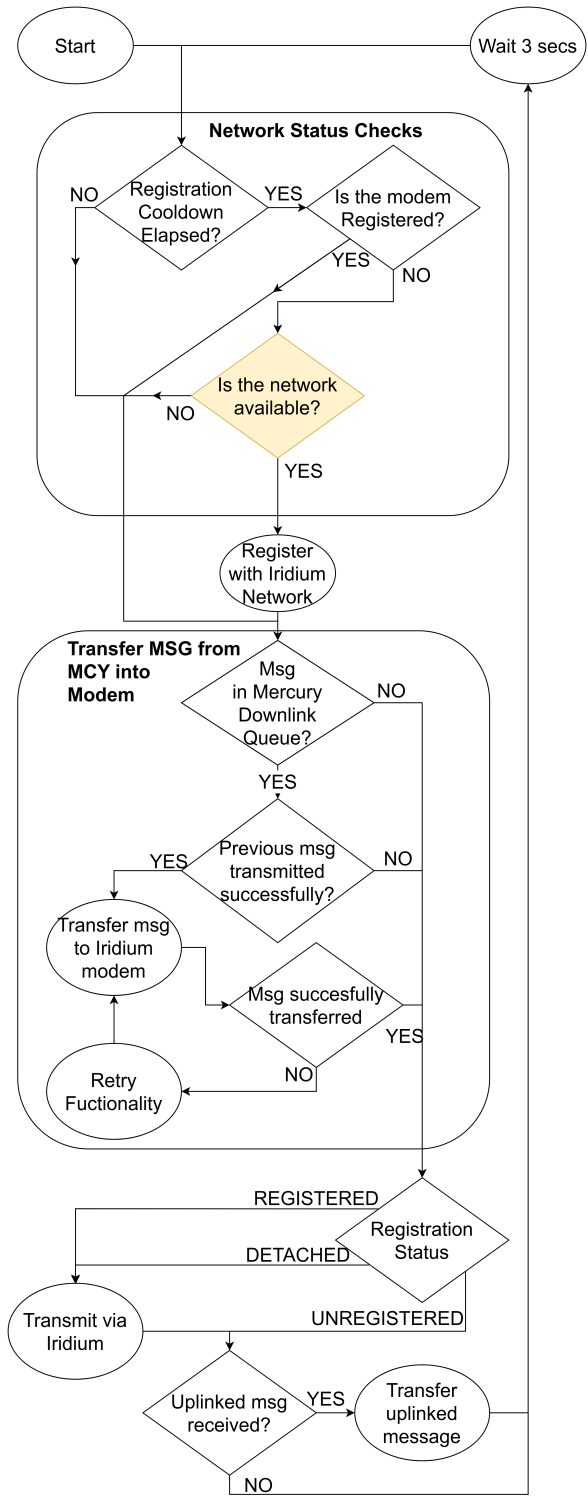

**Figure 10: Blind RxTx Thread Flowchart**

**RSSI Mapping:** In order to determine the likelihood of contact being made with the network, given a UT's orbital position, it may be insufficient to assume the conical model of the beam (Figure 1). To determine the degree to which the conical model can be expanded at a given altitude, it is necessary to map the signal strength received from the network satellites while in orbit. The Iridium modem provides a coarse measurement of this signal strength via the Relative Signal Strength Indicator (RSSI) measurement. This experiment logs the RSSI value reported by the modem, providing a coarse quantization of the strength of the last 'ring-channel' pilot signal detected by the Iridium modem every second.

However, as the RSSI number indicates the signal strength relative to a fixed but arbitrary value, and since it is possible that the iridium modem and antenna degrade over the course of the mission lifetime, RSSI data points taken over a beam traversal will be normalized using the mid-point of the traversal in order to compare values between experiments. A major assumption made here is that the signal strength of a beam is both axially symmetric, and to first order, is identical between each satellite, meaning that for any given $d_{alt,\varphi}$, the logged RSSI values will fit to the same curve for any Iridium satellite and a given $\varphi$.

At each logged event the following data is recorded: the GNSS derived time and position of SpIRIT; the temperature of the modem and antenna; the RSSI value (6 level quantization); the Iridium satellite transmitting the signal; which cell spot of the Iridium beam is transmitting the ring-channel; and the X,Y,Z ECI position of the Iridium satellite.

**Blind Receive and/or Transmit:** The final experiment threads are designed to support unoptimized communication attempts using the Iridium modem.

Rather than using the characterized parameters of the constellation to predict which beam traversals will have a higher likelihood of a successful transmission, the blind receive and/or transmit (BRxTx) threads will continuously attempt to register and transmit/receive to/from the network, foregoing any minimization of the resource utilization and operational opportunity costs of doing so.

The flowchart of the BRxTx threads is shown in Figure 10. This thread attempts to: register with the Iridium network, provided the network conditions are acceptable; ingest a message from the controlling application into Mercury and then load it into the Iridium modem. Finally, if the registration was successful, a message is transmitted via the Iridium network. If the Tx functionality of the thread is not activated the step of transferring a message from Mercury into the modem is skipped, and a 0 byte message is transmitted via the network, this is referred to as a 'mail-box check' operation by the Iridium network and will deliver any waiting messages to the modem..

**Optimized Communication Attempts:** Once data from the characterization experiments have been analyzed, derived characterization parameters will be used to inform both resource utilization, and timeliness optimal communication attempt strategies. Due to schedule pressure, these optimal strategies have not been implemented in Mercury directly, but instead will leverage the operational scripting capabilities of SpIRIT, and use the direct Iridium modem commanding functionality built into Mercury.

**EARLY ON-ORBIT RESULTS**

The SpIRIT mission was launched into a 510km Polar sun-synchronous orbit (SSO) on December 1st, 2023. Since that time, commissioning activities of the various payloads has been ongoing with promising results from all University of Melbourne payloads, a companion paper [18] describes the early commissioning results of other payloads and the HERMES instrument.

The Mercury payload has demonstrated nominal behavior of all electrical and COTS systems, with temperature monitoring provided by Mercury being used during the early operational phases throughout December 2023 to assess operational stability of other payloads.

Further commissioning activities of the Mercury payload are on-going, with a number of compounding reasons for the protracted nature of the commissioning. These are listed below in increasing level of impact.

1. Operational time and resources are shared among the various payloads, with the primary payload (the HERMES instrument) receiving the bulk of operational time by design, in order to carry out commissioning activities related to its operations.

2. As the S-Band radio on SpIRIT has yet to be fully commissioned [18], data down-link is only via the low data rate primary UHF link, limiting both the quantity of Mercury specific data available, and decreasing the cadence of possible data down-link due to load-balancing of the link between payloads and between payload and spacecraft data.

3. Limited contact with the GNSS constellations.

4. Unexpected behavior in interactions between the Iridium network and Mercury

**GNSS Results:** A GNSS time and position lock was first acquired on January 22$^{nd}$, 2024, with intermittent success since then in maintaining a position lock throughout spacecraft operations. Figure 11 shows the location of the Mercury payload (SpIRIT nanosatellite) on-orbit during attempts to receive time and position fixes with either of the two Mercury GNSS modules during the period Jan 22$^{nd}$ → April 27$^{th}$ 2024. Figure 11 shows successful time fixes in orange, and successful position fixes in green. Those locations where neither a time or position fix was attained are shown in red. Two major areas, centered on the South Atlantic Anomaly, and Australia, show no recorded samples. These areas were excluded from commissioning operations for the Mercury payload, the former area was excluded to minimize potential radiation damage accumulated while the payload was on, while the latter was excluded in order to avoid interference during UHF ground-station passes throughout the commissioning period. Outside of these two regions, there seems to be little correlation with geographic location, with position fixes being obtained at all latitudes, and at least time fixes across most longitudes.

Further analysis and amelioration efforts are ongoing.

**Iridium Network Contact Results:** To date, throughout commissioning activities the Mercury payload has not been able to transmit a message via the Iridium network. Efforts have been ongoing to ascertain the root cause of this.

During early commissioning efforts, the blind BRxTx thread was used to attempt to send a message via the network, while also periodically polling the registration status internally held by the Iridium modem. This internal record of the modem's registration status should indicate if the modem was able to make contact with the network, as a transition from the Unregistered state → a Detached state is believed to only occur if the transmission with the network completed (although it is possible the registration attempt was rejected by the network). Due to the polled sampling of this status, an event transition may occur any time during the sampling period. During intermittent testing over available operational windows, the transitions shown in Table 3 were recorded.

From Figure 12 it is clear that the suspected registration events were independent of GPS location knowledge. It is necessary (though not necessarily sufficient) for a registration attempt made by the Iridium modem to provide the network with the UT's current location. Given the lack of position fix during these suspected events, it is unlikely that the position by the UT (derived from the Mercury GNSS modules) would be accepted by the network, and consequently a queued message would not be transmitted (refused by the network). Despite this possible explanation of lack of successful message transmission, it does not explain the sparsity of potential registration events throughout testing.

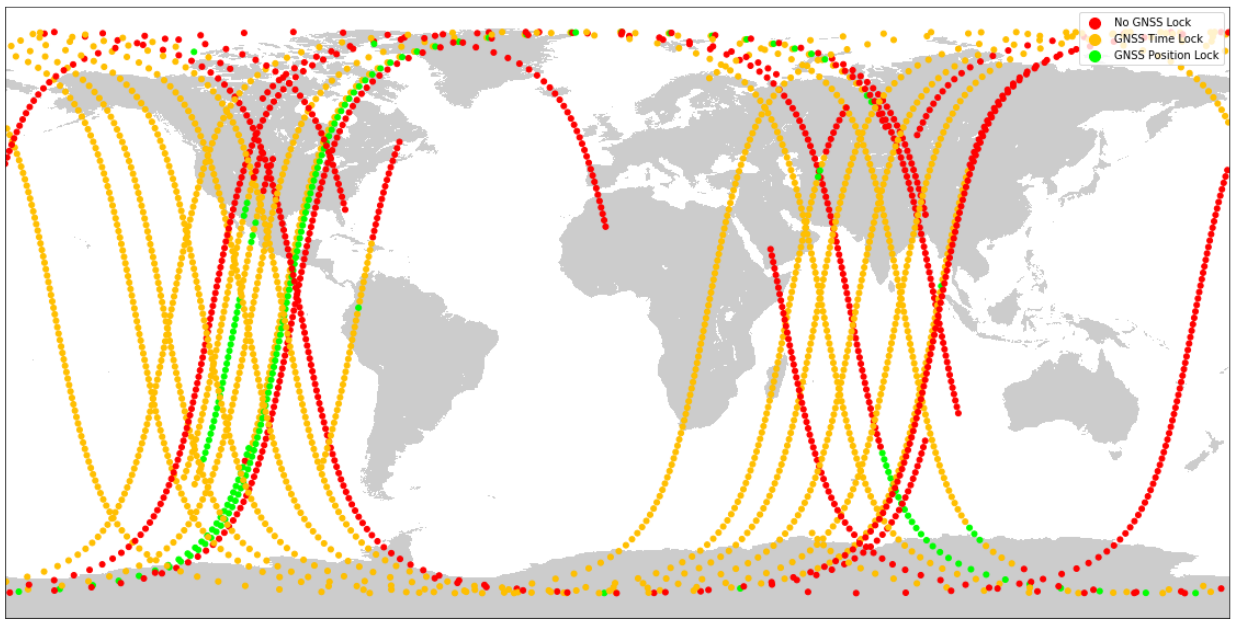

**Figure 11: Location of Mercury on-orbit GNSS time and position locks during early commissioning operations; Red - TLE based spacecraft location with no GPS information, Orange – TLE based spacecraft location where sufficient GPS L1 signals were received to discipline the GNSS module oscillator; Green – location with sufficient L1 signals to maintain position updates.**

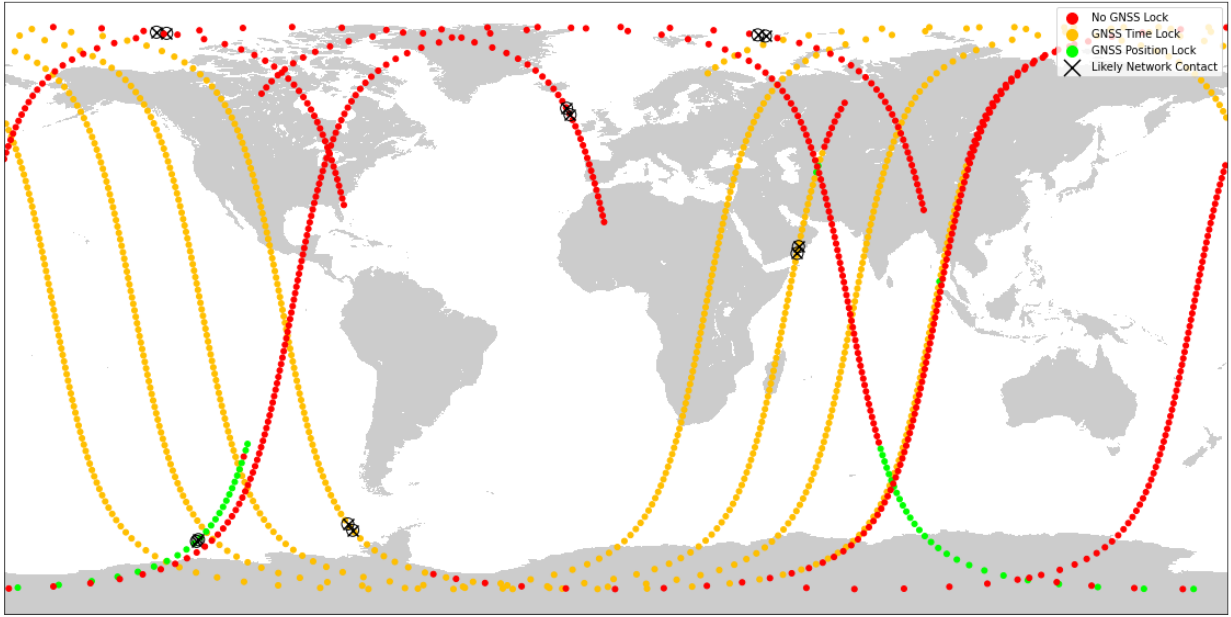

**Figure 12: Location of likely Mercury registration events on-orbit during early commissioning operations; Red - TLE based spacecraft location with no GPS information, Orange – TLE based spacecraft location where sufficient GPS L1 signals were received to discipline the GNSS module oscillator; Green – location with sufficient L1 signals to maintain position updates; Black – Location of likely registration attempts.**

After further testing both on-orbit and ground, the suspected cause is the network availability guard (highlighted yellow in Figure 10). This guard represents a fundamental flaw in the logic of the BRxTx thread flow, and introduces a potential race condition during its operation.

The Iridium modem periodically searches for a ring-channel signal transmitted by network satellites, one per cell spot. If a signal is not found, the modem employs an unpublished algorithm to gradually increase the intervals between successive searches, with a reported maximum search interval of 120s [19]. This search back-off algorithm is designed to reduce the power consumption of the modem in times of low signal quality. The result of this ring-channel search is used by the UT to both report the network signal quality (RSSI value) and report signal availability (equivalent to RSSI > 0).

**Table 3: Likely Mercury Iridium Registration Events**

| Prior Sample | | Post Sample | | |
|---|---|---|---|---|
| Date Time [UTC] | Registration Status | Date Time [UTC] | Registration Status | Sample Period [sec] |
| 2024-02-10 04:13:56 | Unregistered | 2024-02-10 04:14:27 | Detached | 31 |
| 2024-02-11 06:33:18 | Unregistered | 2024-02-11 06:33:50 | Detached | 32 |
| 2024-04-25 16:53:59 | Unregistered | 2024-04-25 16:54:08 | Detached | 9 |
| 2024-04-26 03:14:40 | Unregistered | 2024-04-26 03:14:49 | Detached | 9 |
| 2024-04-26 15:53:13 | Unregistered | 2024-04-26 15:53:21 | Detached | 8 |
| 2024-04-26 22:05:54 | Unregistered | 2024-04-26 22:06:25 | Detached | 31 |

After running for any significant length of time (several minutes) the likelihood of a ring-channel search (once every 120s) coinciding with a beam traversal (maximum duration of ~47s) is qualitatively low. The exact chance of these events coinciding will depend on the relative: inclination, altitude, orbital precession, RAAN, of the UT orbit to the network orbits.

In the case of the BRxTx thread, therefore, the likelihood of progressing beyond the 'Signal Available' guard is low, potentially preventing the thread from ever registering and transmitting a message. Critically, the 'Send Registration' command of the Iridium modem immediately forces a ring-channel search. Consequently, placing the 'Send Registration' command behind the guard is both unnecessary and counter-productive.

Unfortunately this guard has been employed in the Registration Timing, Message Transmission Timing, and as already discussed, the Blind Receive and Transmit threads. Similarly, while not explicitly included in the RSSI mapping thread, the back-off algorithm will also affect this experiment, since a positive RSSI value can only be recorded when a ring-channel search has successfully completed, when this search occurs is governed by the back-off algorithm.

Continuing Mercury commissioning activities are focusing on determining methods to force a reset of the search back-off algorithm increasing the likelihood of a search event coinciding with a beam traversal, as well as experimentally characterizing the back-off algorithm to ascertain an optimal search reset period. Early results show some promise.

Figure 13 and Figure 14 show ring-channel signal detections relative to Iridium network satellite traversals during the relevant period. Figure 13 employed power cycling the entire Iridium modem to achieve a reset of the back-off algorithm, while Figure 14 used the inbuilt ability of the Iridium modem to enable/disable the modem's radio, triggering a reset of the back-off algorithm. These two data sets represent a binary analogue of the RSSI mapping experiment.

It should be noted, however, that in both Figure 13 and Figure 14, the beam traversals are determined from TLE propagation and so include inherent uncertainty in the timing of the start and end of the beam traversals. Nevertheless, it is interesting to note that for those traversals where a ring-channel signal was detected, the resulting detection pattern looks very similar; the signal detection occurs in the $2^{nd}$ half of the traversal, with a brief loss of signal.

The implications of this detection pattern on the planned experiments, as well as the non-detections during earlier beam traversals in Figure 14 are currently under investigation.

**ONGOING INVESTIGATIONS**

Further to continuing the above binary coverage mapping strategy to investigate both methods for and impacts of strategies for resetting the ring-channel search back-off algorithm, efforts are underway to empirically determine the structure of the back-off algorithm. Once the algorithm is better understood, periodic resets can be triggered at defined intervals to minimize both the effects of the back-off algorithm on the outlined experiments, as well as the effect of the loss of signal from the reset itself.

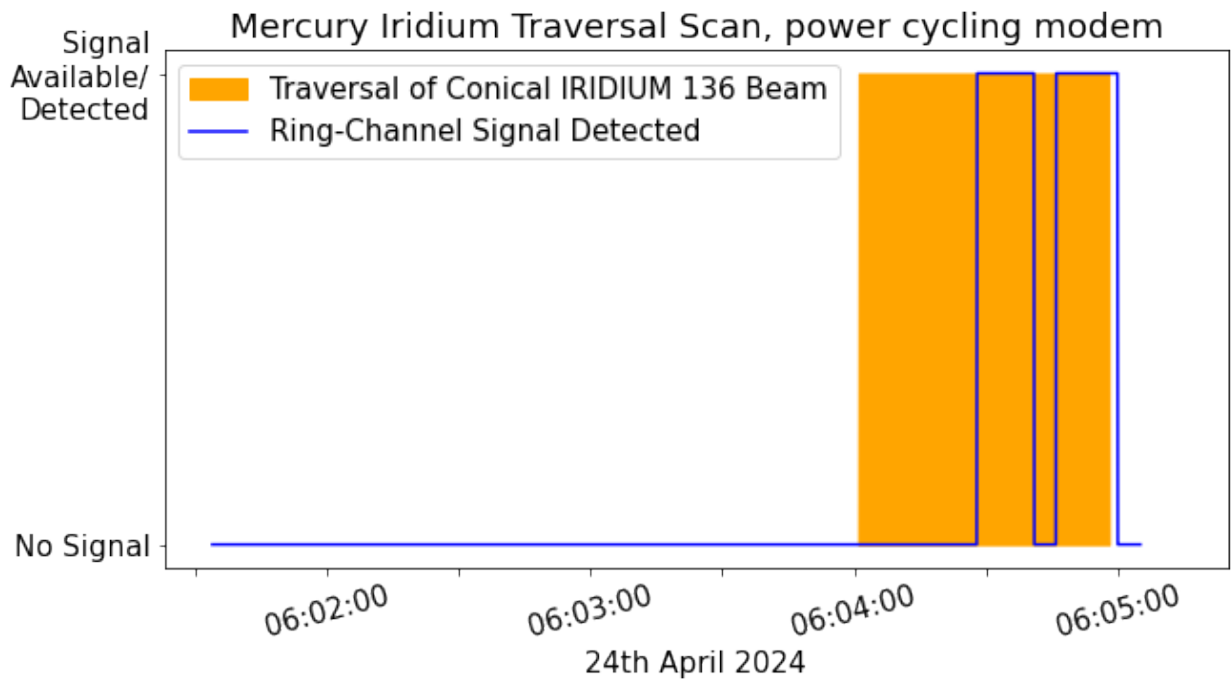

**Figure 13:** Iridium network detection events by the Mercury Payload. Ring-channel search reset by power cycling the modem.

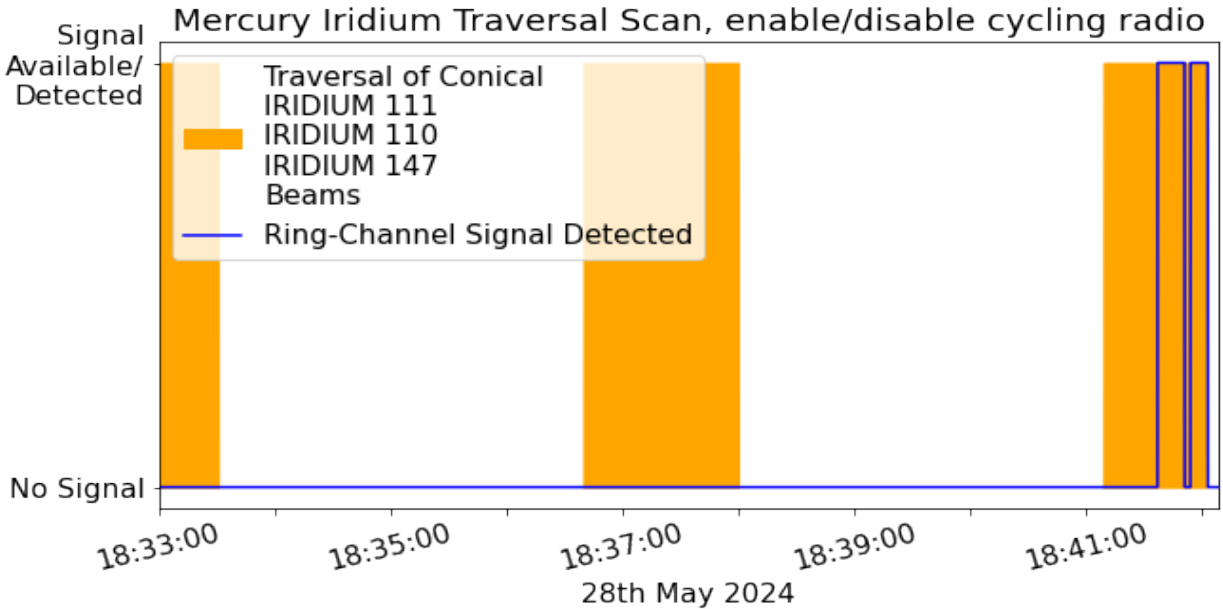

**Figure 14: Iridium network detection events by the Mercury Payload. Ring-channel search reset by disabling and re-enabling the modem radio.**

Additionally, efforts are underway to reimplement the BRxTx thread functionality using SpIRIT's operational scripting capabilities, without the Signal Available guard.

Finally, further data sets of GNSS fixes are being collected to fully understand any underlying reasons for the lack of continuous GNSS fix.

**LESSONS LEARNT**

Although significant work remains to fully commission the Mercury payload, and achieve an operational state sufficient to reliably gather experimental data suitable for characterizing the Iridium network, there are a number of valuable lessons learnt from operating the Mercury payload at this early stage.

The requirements governing the selection of GNSS modules were not fully elicited. Although the HERMES instrument required high-precision timing information, the effect of this on the provision of position fixes for the Mercury payload were not fully explored. It is possible that due to the high-precision timing capability, the Venus modules must wait for its internal algorithms to reduce the Dilution of Precision (DoP) value to fall below a particular threshold, either through more samples, or a greater number of visible GNSS satellites to be visible before a timing and position fix will be output. Given that the position fix required for supply to the Iridium network is of significantly lower precision requirements, a fix with a 'looser' DoP may have been acceptable.

A more thorough understanding of the mechanism and implications of the ring-channel search procedure within the Iridium modem during payload development would likely have prevented the unnecessary 'Signal Availability' guard being instituted. This is an unfortunate circumstance which given either more detailed information regarding the workings of the modem (an impossible ask given the proprietary nature of the modem) during development, or more realistically, more thorough testing on-ground during development could have been avoided. Although the testing of a UT on the ground but under orbital conditions, would be extremely challenging, (direct modification of the transmitted/received signals at the modem using up/down mixing to simulate doppler shifts due to orbital motion and RF switching to simulate intermittent coverage at the desired altitude), it would be extremely beneficial to uncover incorrect assumptions made during the design stage. There are of course still limitations to such a simulated testing apparatus, particularly the inability to simulate any latitude/longitude dependent behavior.

## CONCLUSION

The Mercury payload, developed by the Melbourne Space Laboratory at the University of Melbourne, Australia was designed and fabricated for the SpIRIT nanosatellite mission, in order to provide a test bed for the characterization of constellation parameters affecting the utilization of SATCOM constellations by nanosatellites. While early on-orbit results have not yet yielded direct communication with the Iridium network, ongoing commissioning activities are yielding a wealth of data providing refinements to the model to be used for optimizing communication attempts with the Iridium network.